\documentclass[12pt,a4paper,notitlepage]{article}
\usepackage[a4paper]{geometry}
\usepackage[utf8]{inputenc}
\usepackage[english]{babel}
\usepackage{lipsum}
\usepackage[round]{natbib} 
\usepackage{amsmath, amssymb, amsfonts, amsthm, mathtools}
\usepackage{multicol}
\usepackage{microtype} 
\usepackage{graphicx}
\graphicspath{ {./pics/} {./eps/}}
\usepackage{epsfig}
\usepackage{epstopdf}
\usepackage[utf8]{inputenc} 
\usepackage[T1]{fontenc}    
\usepackage{hyperref}       
\usepackage{url}            
\usepackage{booktabs}       
\usepackage{amsfonts}        
\usepackage{nicefrac}       
\usepackage{lipsum}

\usepackage[T1]{fontenc}
\usepackage{lmodern}

\title{Stock Price Forecasting and Hypothesis Testing Using Neural Networks}

\author{
  Kerda Varaku \\
  Department of Economics\\
  Rice University\\
  \texttt{kerda.varaku@rice.edu} \\
  \date{}
}

\begin{document} 
\maketitle
\begin{abstract}

In this work we use Recurrent Neural Networks and Multilayer Perceptrons, to predict NYSE, NASDAQ and AMEX stock prices from historical data. We experiment with different architectures and compare data normalization techniques. Then, we leverage those findings to question the efficient-market hypothesis through a formal statistical test.
\end{abstract}

\section{Introduction}
Predicting stock returns is a task that has attracted a lot of attention among investors, financial managers and advisors but also among academics. This is important to finance practitioners to best allocate their assets and to academics to build better and more accurate asset pricing models. Moreover, predicting stock returns gives crucial implications about market efficiency. Interest in stock price movements dates back to \cite{kendall1953analysis} where it was noticed that stock prices seemed to move randomly over time. This is what is called ``the random walk hypothesis". This hypothesis is consistent with the efficient-market hypothesis which views prices as a function of information. The efficient-market hypothesis states that current stock prices reflect all the available information and if there are movements in the next periods, they comes as a result of a release of new information or as a result of random shocks. This means that it is impossible to predict future stock prices using past information. This work uses Artificial Neural Networks (hereafter ANNs) to question efficient market hypothesis by attempting to predict future individual stock prices using historical data.

\subsection{Related Work}
The efficient markets theory was first proposed by the French mathematician Louis Bachelier in 1900 (see \cite{bachelier1900theorie}) but it started to draw a lot of attention only by the 1960’s. There were a huge number of studies which analyzed whether this hypothesis was true. The first variables used in predicting future movements were past prices, and later other predictive variables such as interest rates, default spreads, dividend yield, the book-to-market ratio, and the earnings-price ratio. (e.g \cite{fama1977asset},  \cite{fama1988dividend}, \cite{rozeff1984dividend}, \cite{shiller1984stock}, \cite{flood1986evaluation}, \cite{campbell1988dividend}). The last three ratios have received the most interest in the literature. The hypothesis that dividend yields forecast stock returns is even older (for example, \cite{dow1920scientific} and \cite{ball1978anomalies}). All three of these ratios have prices in the denominator, thus, high rations imply that the stock is undervalued, which in turn suggests high subsequent returns. This contradicts the view that returns (and thus prices) can not be predicted using past information. Indeed, studies such as \cite{fama1988dividend} showed that these ratios are positively correlated with subsequent returns. 

The interest in the topic continued also during the 1990s
and early 2000s where in addition to the above mentioned predictive variables, many authors like \cite{lamont1998earnings}, \cite{baker2000equity}, \cite{lettau2001consumption}, used financing activity, consumption/wealth relation and valuations of high and low beta stocks. Setting aside the difference in predictive variables, model specifications or different ways to correct for errors and/or biases, what all these studies have in common is the usage of traditional approaches and methods for estimation and predictions. \cite{fama1988dividend} use a linear least-squares model to predict the monthly NYSE returns from dividend yields and they find a $t$-statistics between $2.20$ and $3.21$. In line with the criticism that the method used here is biased (e.g  \cite{stambaugh1985bias} and \cite{mankiw1986we}), \cite{nelson1993predictable} replicate the study, correcting the bias by using bootstrap simulations. They find $p$-values between $0.03$ and $0.33$. On the other hand, \cite{stambaugh1999predictive} performs the estimation assuming that dividend yields follow a first-order autoregressive (AR1) process. He regresses the NYSE returns during 1952-1996 on dividend yields and the $p$-value reported for the one-sided test is $0.15$.

Although the focus has been on predictive variables as the ratios mentioned above, stock prices are also very sensitive to social and political events, monetary policy, interest rates, and many more macro economic variables. Public news announcements or periods of no public information release also significantly affect fluctuations in stock prices (e.g \cite{chen2000extensions} and \cite{chaudhuri2004stock}). The multivariate vector autoregression (VAR) model, a generalization to univariate autoregressive (AR) model, has also been widely used in the literature. This is a stochastic process model, treating all variables as potentially endogenous, and it is used to capture the linear interdependencies among multiple time series.

Nevertheless, given the chaotic, extremely volatile and nonparametric nature of stock prices, also these methods are questioned when it comes to obtaining reliable results. If stock prices do not follow random walks, what processes do they follow? When it comes to observing patterns, various studies have shown that stock returns exhibit reversal at weekly and 3-5 year intervals, and drift over 12-month periods (\cite{bondt1985does}, \cite{lo1990contrarian}, \cite{jegadeesh1993returns}). Several models have been proposed to capture the predictability of stock returns using the approaches mentioned above. \cite{grossman1996equilibrium} evaluates some popular models using a Kalman Filter technique and finds that they have serious flaws. Popular models are usually too restrictive, they fail to perform well in empirical tests and have even worse performance in out of sample tests.

Despite the difficulties of the task, different estimation models, different ways of correcting for biases, there is a consensus among financial economists nowadays that stock returns contain a significant predictable component based on in-sample tests (Campbell 2000). Nevertheless, in the out of sample forecasting tasks, the predictive ability is pretty low (\cite{bossaerts1999implementing} and \cite{goyal2003predicting}, \cite{welch2008goyal}). This is mainly related to the dynamic, and complex nature of the markets. Considering the difficulty and the flaws of the traditional estimation and forecast approaches, ANNs are now widely used in finance and economics and have received special focus especially on the task of stock price forecast. They have achieved high accuracy even in the out of sample, or test sets. ANNs have become an attractive tool for predicting stock market trends mostly because they can predict any non-linear dynamic system to any accuracy provided some mild conditions are met.

In one of the earliest studies, \cite{kimoto1990stock} used several learning algorithms and prediction methods for the Tokyo stock exchange prices index (TOPIX) prediction system. Their system used modular neural networks to learn the relationships among various factors. \cite{kamijo1990stock} used recurrent neural networks for stock price pattern recognition and \cite{ahmadi1990testability} used a neural network to study the arbitrage in the stock market. \cite{yoon1991predicting} also performed predictions using neural networks. Some were focused on forecasting the stock index futures market.
\cite{trippi1992trading} and \cite{choi1995trading}
predicted the daily direction of change in the SP 500 index futures using ANNs. \cite{duke1993neural} tried to predict the daily predictions of the German government
bond futures. Even though many of them did not bear
outstanding prediction accuracy, many others have shown
that ANN approach can outperform conventional methods
(eg. \cite{yao1995forecasting}; \cite{van1996application}; \cite{fernandez2000profitability}).

Most of the studies that use ANN to predict movements in stock prices use high frequency data, i.e hourly or daily data. \cite{martinez2009artificial} analyze intra-day price movements and predict the best times to trade and make profits within a day. Other studies like \cite{senol2008stock} try to predict the direction of the movements rather than how much will the stock move, where the output is a categorical variable (goes up, goes down or stays the same). \cite{yao1995forecasting} also use daily data to forecast the Kuala Lumpur Stock Exchange (KLSE) index. Many other studies focus on predicting indexes rather than individual stock prices such as \cite{wanjawa2014ann}, \cite{sheta2015comparison}.

\subsection{Contributions}
This work has the following novel contributions in stock price forecasting: 1) it does not make assumptions related to the functional form of the model, 2) it aims at predicting individual stock prices rather than predict price indexes, 3) it tries, given historical data, to predict one quarter or one month ahead rather than dealing with intraday or next day predictions, 4) it only uses as input the historical stock price of each firm and 5) it also offers a test for the relevance of network inputs in the prediction task. 

\section{Data and Methodology}
\subsection{Data and statistics}

\begin{figure}
\centering
  \includegraphics[width=0.47\linewidth]{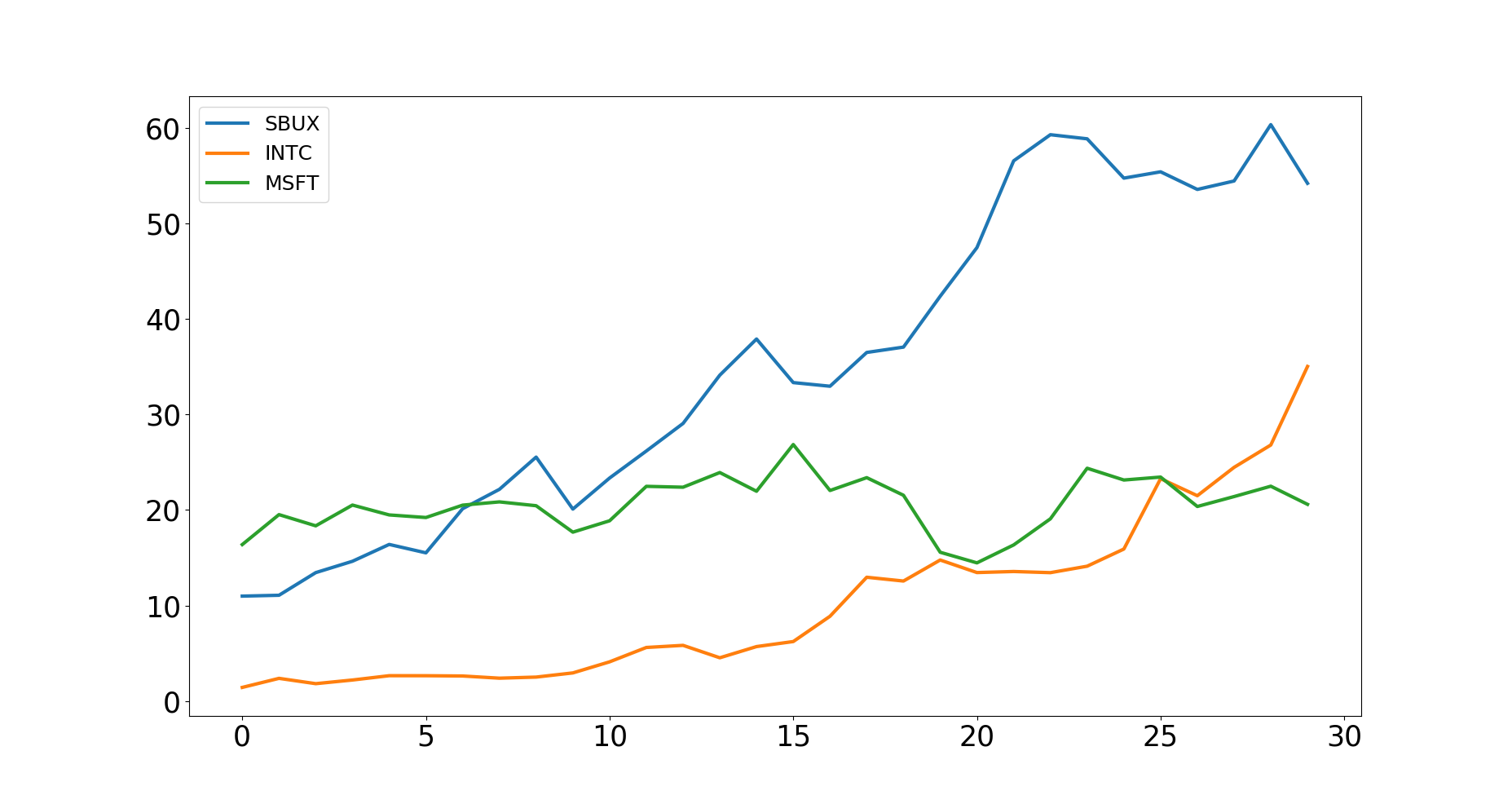}
  \includegraphics[width=0.47\linewidth]{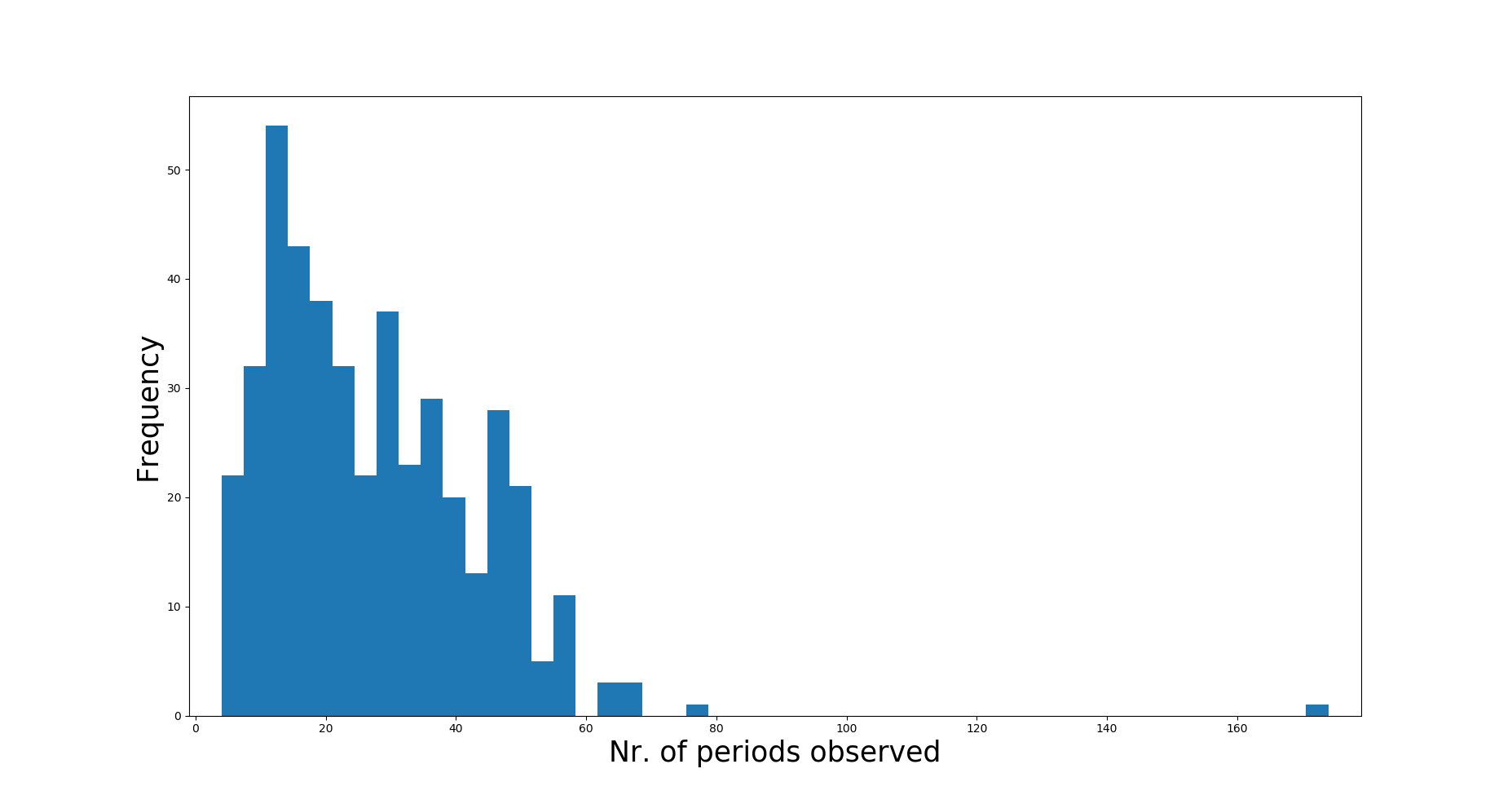}
\caption{Left: Sample stock prices over time, Right:Frequency of number of observed years }
\label{Ffig}
\end{figure}

Historical data of prices and dividend yields are taken
from Quandl (\cite{QuandlDT}) with data that ranges from 1980 to 2017. Quarterly data on prices, i.e the closing price at the end of each quarter is used and later on, monthly frequency data, i.e the closing price at the end of each month, is used. Quandl provides stock prices data for stocks in NYSE, NASDAQ and AMEX. The number of companies used in the study is 439. On the left, Fig. \ref{Ffig} shows the time series of three sample stock prices. Some prices have a very steep slope of the time trend, for some the slope of the trend is lower, and some of them appear more stationary and oscillate around some average price value during time, even though the variance seems to change. These patters highlight the highly non stationary nature of the prices. 

Each of the companies is observed for a different number of quarters. The shortest time series is restricted to 16 quarters. The longest one consists of 175 quarterly observations for different companies in the sample. Monthly time series is also restricted to have a length of at least 16. Fig. \ref{Ffig} on the right shows the histogram of the number of years observed for 439 firms in the sample.

Different models require different ways to deal with the
time series of different length and different processing of the data. We consider two models in order to forecast future quarterly stock prices: Recurrent Neural Network (hereafter RNN) and Multilayer Perceptron (hereafter MLP). We also consider an MLP with monthly frequency data and for the latter case, we also offer a test on whether past changes in stock prices have any effect on future prices (Section \ref{testsection}). The next subsections will explain in detail how the data is processed for each of the models considered. The MLP with monthly data is similar to the one with quarterly data. 

In all the following sections, we have used TensorFlow library (\cite{tensorflow2015-whitepaper}) in Python and NVIDIA® GeForce® 940MX GPU. 

\subsection{Recurrent Neural Network}
In this subsection we discuss how the data is processed in order to be used in an RNN (see \cite{graves2013speech}). Later on, the RNN architecture is discussed, and lastly, how learning is conducted. 

Given the time series nature of the data, a natural way for prediction is using an RNN. After having a choice of modeling, the data must also be processed accordingly. The data in Quandl is obtained in the form of a data-frame and then it is reconstructed as a dictionary, with stock name as key and the associated matrix, $M_n$, as value. For stock $n$ , where $n \in \{1, . . . , N \}$, the matrix $M_n$ is a $T_n \times 3$ matrix where $T_n$ is the time period for which data is available on stock $n$ and the three columns correspond to values of time, past prices and current prices respectively. The second column is a shift of the third column one row above. The first observation is lost, since there is no history of price to predict the price value of the first observation. We need to store the data as a $N \times T \times 3$ matrix where $T = \max \{T_1 ,...,T_N \}$. For this purpose, we need all the firms to have the same time dimension. Thus, we replace the non available data (the missing values) of each matrix $M_n$ with zeros. The original matrix $M_n$ and the augmented matrix $\tilde{M}_n$ are shown below:

\noindent\begin{minipage}{.5\linewidth}
\begin{equation}
 M_n=
\begin{bmatrix}
t_1 &  x_0  & x_1\\
t_2  &  x_1 & x_2\\
\vdots & \vdots & \vdots\\
t_{T_n}  &   x_{T_n-1}   & x_{T_n},
\end{bmatrix},
\end{equation}
\end{minipage}%
\begin{minipage}{.5\linewidth}
\begin{equation}
 \tilde{M}_n
\begin{bmatrix}
t_{-(T-T_n)} & 0 & 0\\
\vdots & \vdots & \vdots\\
t_0 & 0 & 0\\
t_1 & x_0 & x_1\\
t_2 & x_1 & x_2\\
\vdots & \vdots & \vdots\\
t_{T_n} & x_{T_n-1} & x_{T_n}
\end{bmatrix}.
\end{equation}
\end{minipage}

\begin{figure}
\centering
\includegraphics[width=0.8\linewidth]{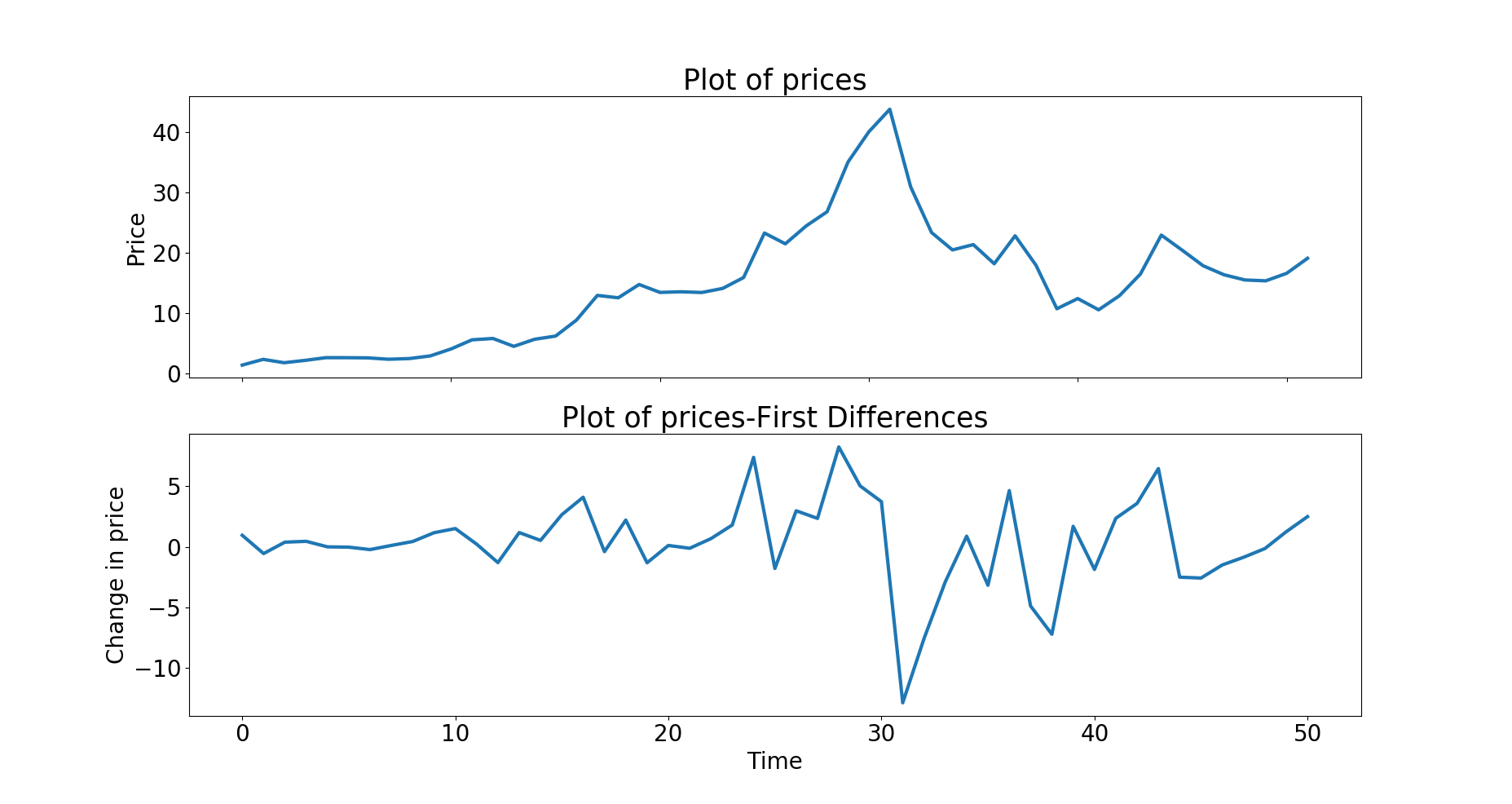}
\caption{Stock price over time for selected sample stock}
\label{Fig3}
\end{figure}

The time column is augmented by the same $\delta_t$ (here it
moves quarterly). The two other columns are added with zeros. For this model we try to use the variables in levels, logs
and also the first difference of the logs. Using the first difference, improves stationary, increasing
the performance after training. Fig. \ref{Fig3} shows the time series of a sample stock in levels and first difference of the logs.

Now the augmented values of $0$ can be interpreted as zero changes in stock prices from one stock to another. A typical RNN architecture for one hidden layer is shown in Fig. \ref{Fig4-5} on the left. In addition to this model, we also use Long Short Term Memory Network (hereafter LSTM) and Gated Recurrent Units (hereafter GRU).

Let $h_{t+1}$ be given by:
\begin{equation}
h_{t+1}=g_h(W_ix_{t+1}+W_rh_t+b_h)
\end{equation}
and the predicted value at time $t$ is given by:
\begin{equation}
\hat{x}_{t+1}=g_y(W_yh_t+b_y).
\label{5}
\end{equation}
$W_i$ , $W_r$ and $W_y$ denote the weights, $b_h$ and $b_y$ denote the biases and $g_h$ and $g_y$ are the non linearities used. The process has an output at every step. First, we need to forward till the end in order to obtain the cost which we denote by $\mathcal{L}$. In this case we use the mean squared error:
\begin{equation}
\mathcal{L}=\frac{1}{N} \sum_{i=1}^N \frac{1}{T} \sum_{t=1}^T (x_{t,i}-\hat{x}_{t,i})^2,
\end{equation}
where $x_{t,i}$ is the actual value of the stock price of company $i$ at time $t$ and $\hat{x}_{t,i}$ is the predicted one given from the Eq. \ref{5}. It is clear that with this formulation, the calculation of the partial derivative of the cost with respect to $W_r$ requires to back-propagate till the beginning in each step. One of the concerns in this case might be vanishing/exploding gradients as well as exponential memory decay over time (see \cite{pascanu2013difficulty}). For this we allow for manipulation of the memory of the system, using LSTM and GRU. They are gated mechanisms or architectures of RNNs that control the information flow through some gates. LSTM is characterized by 3 gates: input, forget and output,
while GRU by 2 gates: reset and update (see \cite{hochreiter1997long} and \cite{cho2014learning}). Using the last two methods increases the memory of the network and they are expected to give more accurate results.

The choice of optimization algorithm for this model is
Adam Optimizer (see \cite{kingma2014adam} and \cite{le2011optimization}). The batch size used is $128$ and we use a
piece wise constant learning rate with boundaries set at
$300000$, $600000$ and $1200000$ and learning rates of $0.00001$, $0.000005$, $0.000001$, $0.0000001$ respectively. A hyperbolic tangent non linearity is used for all the hidden layers, except the last one where no nonlinearity is used. The sample is divided randomly in a train set which consists $67\%$ of the total sample and a test set consisting of $33\%$ of the total sample.

\subsection{Multilayer Perceptron}
In this subsection, the MLP (see \cite{lecun2015deep}) is discussed , including how the data is processed, the architecture of the MLP and how the learning is conducted. 

The input in each period is a vector of past prices, rather than a scalar. Having such a model in mind, the data should be reprocessed accordingly. Now for every firm $n \in {1,..., N }$ the input matrix will be of the form shown in Eq. \ref{Eq7} and the output will be of the form shown in Eq. \ref{Eq7-1}:

\noindent\begin{minipage}{.6\linewidth}
\begin{equation}
  X_n=
\begin{bmatrix}
x_0 & x_1 & \cdots & x_{m-1}\\
x_1 & x_2 & \cdots & x_m\\
\vdots & \vdots & \ddots & \vdots\\
x_{T_n-m-1} & x_{T_n-m-2} & \cdots & x_{T_n-1}
\end{bmatrix},
\label{Eq7}
\end{equation}
\end{minipage}%
\begin{minipage}{.3\linewidth}
\begin{equation}
   y_n=
\begin{bmatrix}
x_m\\
x_{m+1}\\
\vdots\\
x_{T_n}.
\end{bmatrix}.
\label{Eq7-1}
\end{equation}
\end{minipage}

Having this information for all firms, we create $X$ by
stacking all $X_n$ and $y_n$ for all $n \in \{1,..., N \}$, so that the
data is ready to be fed into the network (see Eq. \ref{eq8-1} and Eq. \ref{eq9-1}).

\noindent\begin{minipage}{.5\linewidth}
\begin{equation}
  X=
\begin{bmatrix}
X_1\\
X_2\\
\vdots\\
X_N,
\end{bmatrix}
\label{eq8-1}
\end{equation}
\end{minipage}%
\begin{minipage}{.5\linewidth}
\begin{equation}
  Y=
\begin{bmatrix}
y_1\\
y_2\\
\vdots\\
y_N
\end{bmatrix}.
\label{eq9-1}
\end{equation}
\end{minipage}

\begin{figure}
\centering
\includegraphics[width=0.47\linewidth]{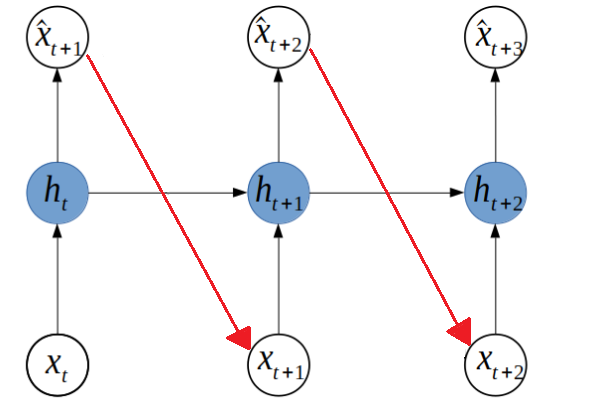}
\includegraphics[width=0.47\linewidth]{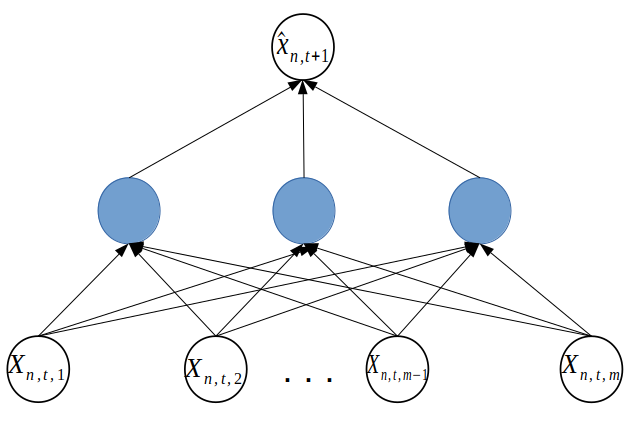}
\caption{Left: RNN Architecture; Right: MLP Architecture. The first subscript denotes the firm, the second subscript denotes the row and the third subscript denotes the column of the matrix in Eq. \ref{Eq7}.}
\label{Fig4-5}
\end{figure}

The architecture of a one layer perceptron is shown in
Fig. \ref{Fig4-5} where in order to predict each period's price of firm $n$
at $t + 1$, the inputs used are $X_{n,t,1}$, $X_{n,t,2}$, ..., $X_{n,t,m}$, for a choice of $m$. Here $X_{n,t,1}$ represents the element in the $t$ -th row and the $1$-st column of the matrix $X_n$. $X_{n,t,2}$ represents the element in the $t$- th row and the $2$- nd column of the matrix $X_n$ and so on.

\begin{figure}
\centering
\includegraphics[width=0.8\linewidth]{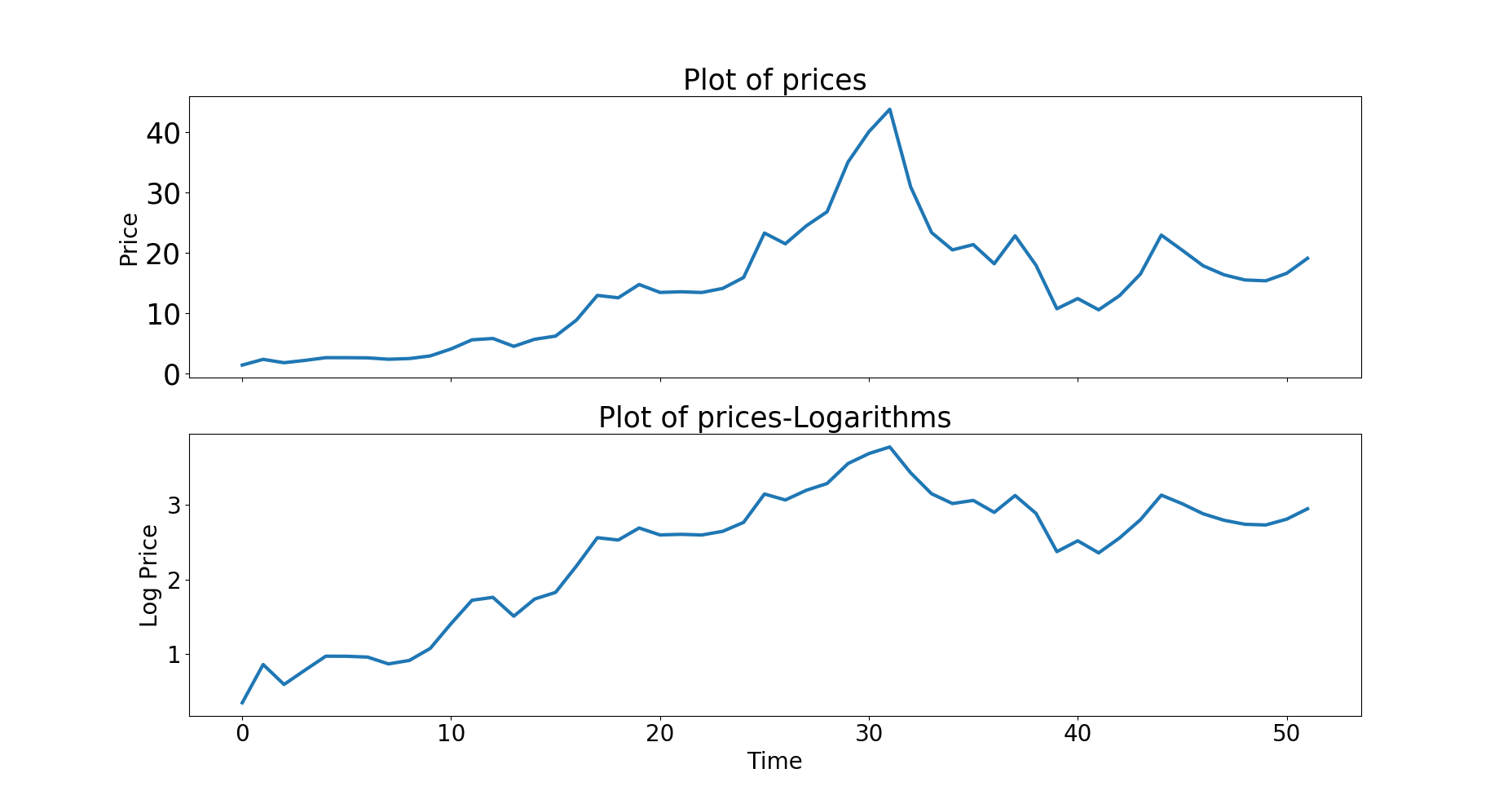}
\caption{Stock price over time for selected sample stock}
\label{Fig6}
\end{figure}

For this model, we use the variables in their logarithmic
form. The time series of price for one of the sample stocks
is shown in Fig. \ref{Fig6} together with its logarithmic transformation. The choice of optimization algorithm for this model is also Adam Optimizer. The loss function chosen is mean squared error, shown in Eq. \ref{Eq11}. The batch size used is $128$ and we use a piecewise constant learning rate with boundaries set at $300000$, $600000$ and $1200000$ and learning rates of $0.00001$, $0.000005$, $0.000001$, $0.0000001$ respectively for each interval.

\begin{equation}
\mathcal{L}=\frac{1}{N} \sum_{n=1}^N \frac{1}{T} \sum_{t=1}^T (y_{n,t}-\hat{y}_{n,t})^2.
\label{Eq11}
\end{equation}

\section{Results}
\subsection{Results from RNN model}
\begin{figure}
\centering
  \includegraphics[width=0.47\linewidth]{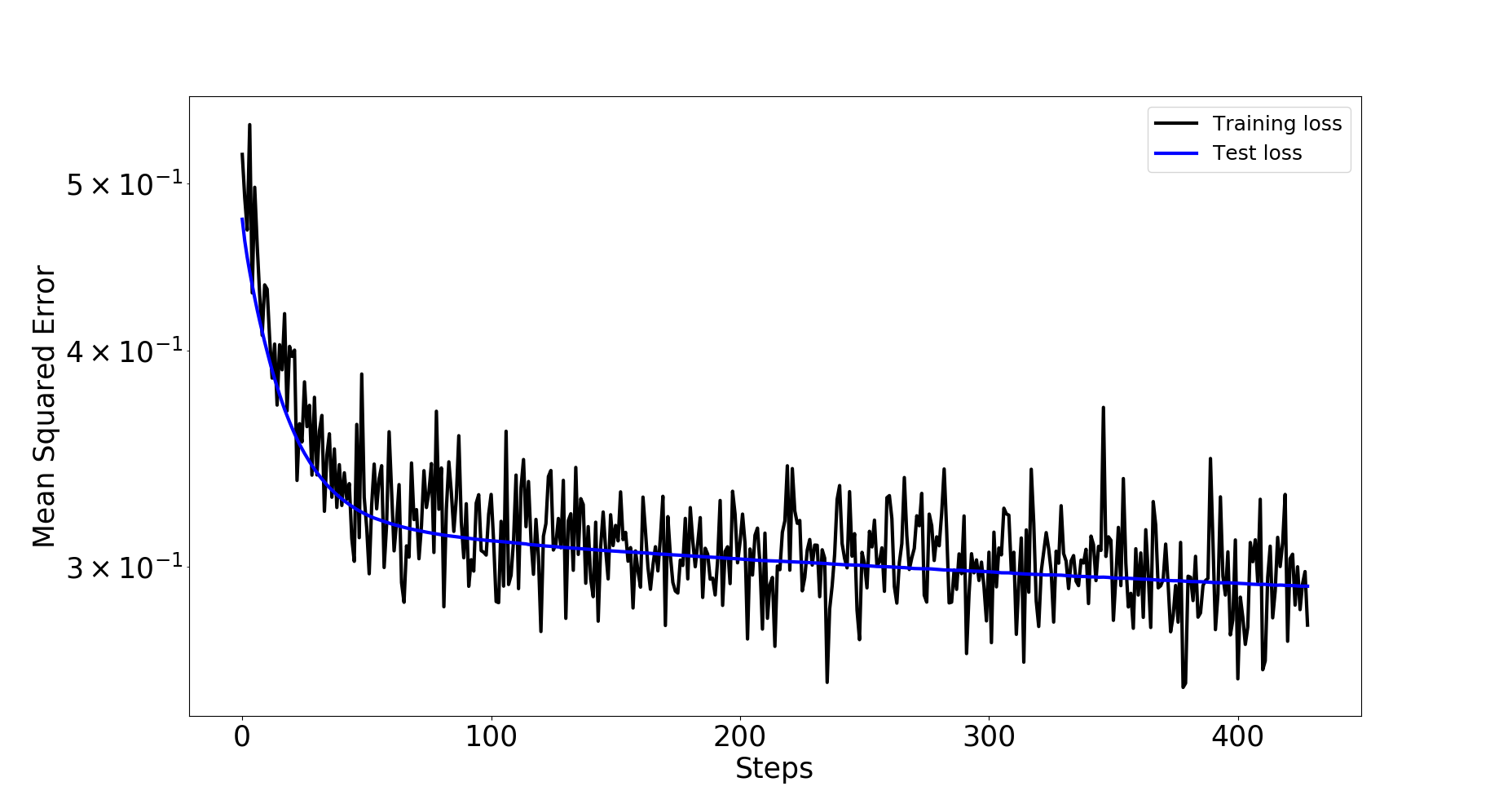}
  \includegraphics[width=0.47\linewidth]{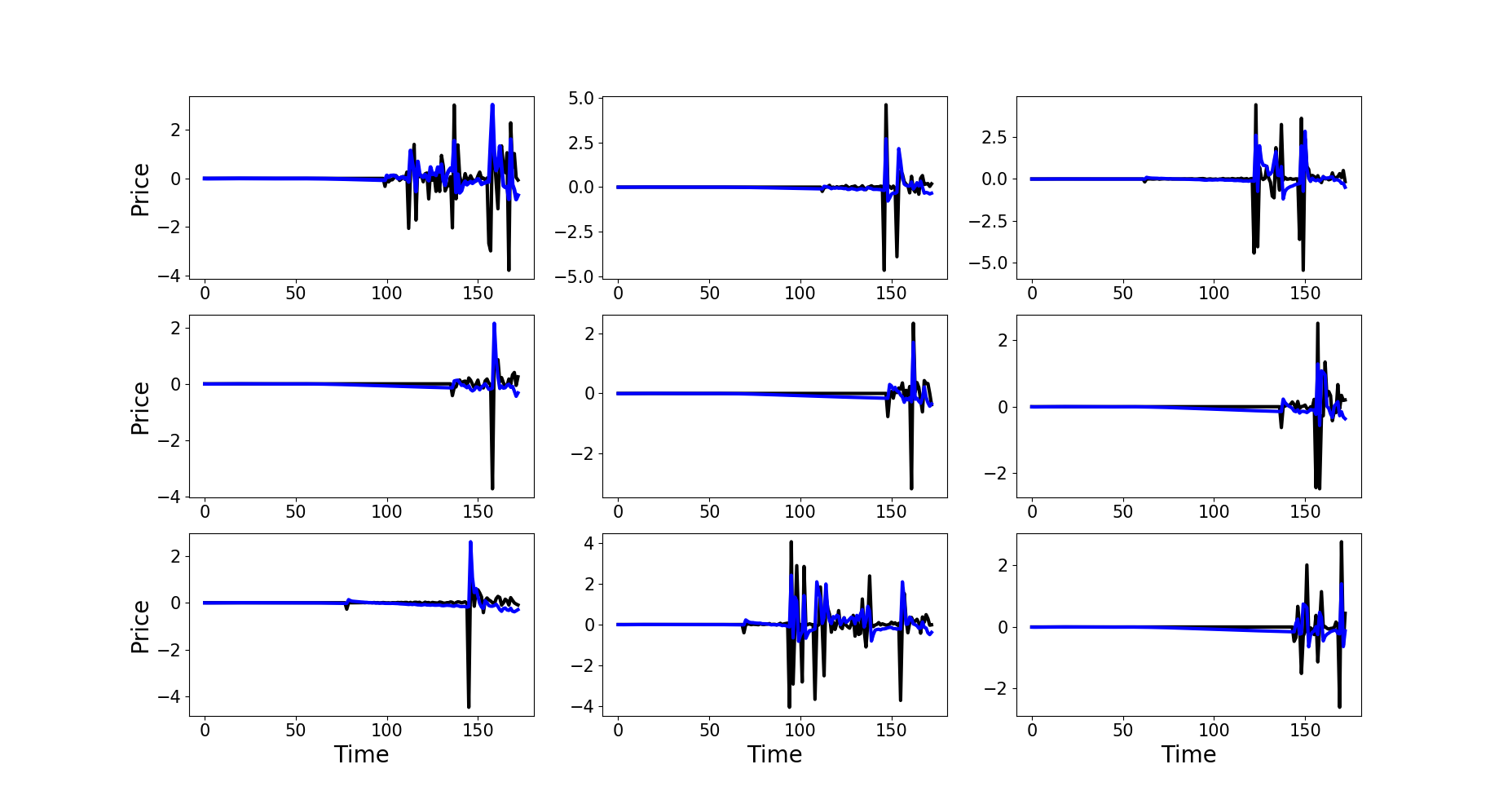}
\caption{Left: Semilogy of the mean squared error loss during training for RNN; Right: Predicted (blue) vs Actual (black) stock prices for random actual-predicted pairs in the test set for RNN}
\label{Fig7-8}
\end{figure}

This subsection discusses the results obtained from the RNN. Fig. \ref{Fig7-8} on the left shows the semilogy of the mean squared error loss during training for every $10$ batches of $128$ samples.
The model used is an RNN with $5$ hidden layers and Basic
cells. Fig.\ref{Fig7-8} on the right shows the predicted values of quarterly prices for random actual-predicted pairs in the test set. The black lines represent the actual time series of prices and the blue line represents the predictions over time. The test loss reached after all the iterations is $0.33$ and the results do not seem very satisfactory. The model fails to predict very high or low deviations from zero.

We repeat the experiment for one and five hidden layers and use basic RNN cell, LSTM cell and GRU cell. Table \ref{Tab1} shows the mean squared error loss in each of the cases.
\begin{table}
\centering
\vspace{2ex}
\begin{tabular}{|l| l| l| l| }\hline
&RNN&LSTM&GRU\\\hline 
$1$ Hidden Layer & $0.2922$ &$0.3076$& $0.3103$\\\hline
$5$ Hidden Layers & $0.3376$& \textbf{0.2628} & $0.2726$\\\hline
\end{tabular}
\caption{Mean Squared Error Comparison}
\label{Tab1}
\end{table}
From these results, we can see that LSTM and GRU achieve the lowest mean squared error with 5 hidden layers. This is because they allow for an increase in the memory of the network compared to the basic RNN cell.

\subsection{Results from the MLP model}
The MLP model results are discussed in the following subsections. The first subsection reports the results from an MLP where the train and test set is split randomly. The next subsection shows the results when train set consists of observations before year 2016 and all the other observations are in the test set.

\subsubsection{Random split of the train and test set}

\begin{figure}
\centering
\includegraphics[width=0.47\linewidth]{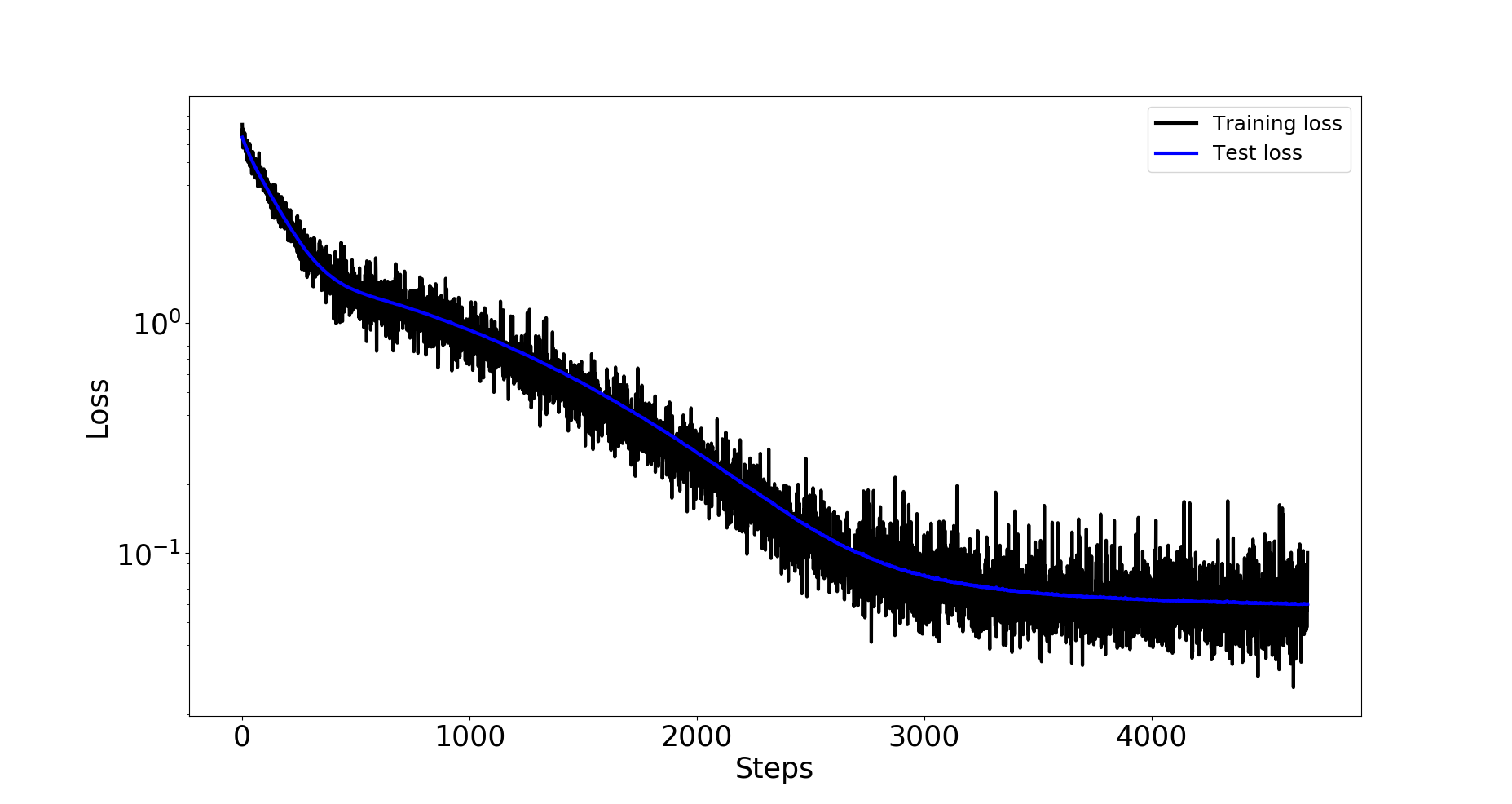}
\includegraphics[width=0.47\linewidth]{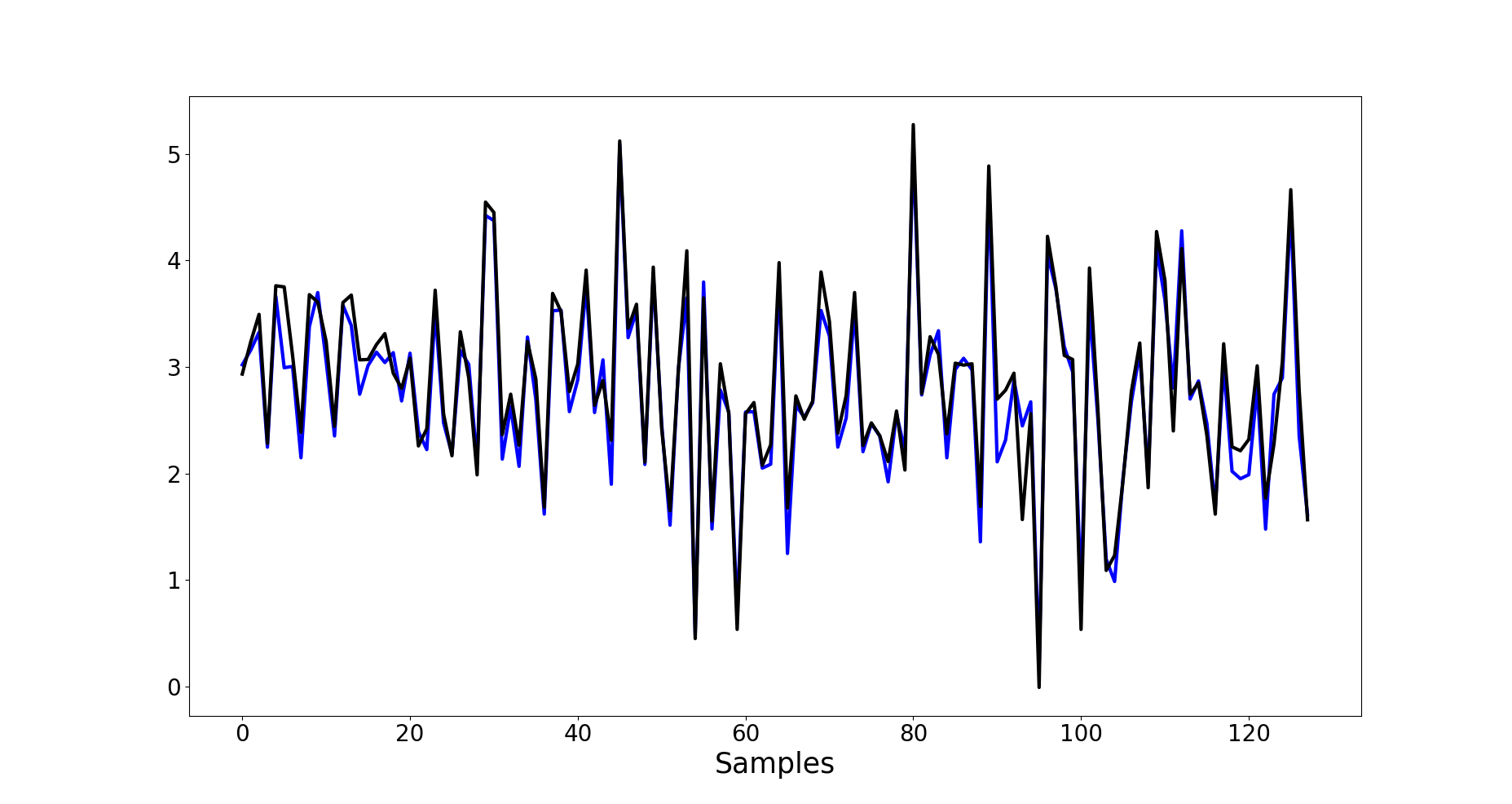}
\caption{Left: Semilogy of the mean squared error loss during training for MLP; Right: Predicted (blue) vs Actual (black) stock prices for random actual-predicted pairs in the test set for MLP with random splitting }
\label{Fig9-10}
\end{figure}

First, we report results from an MLP model with $1$
hidden layer, $x_{n,t}$ and $9$ lags of $x_{n,t}$ to predict the next price $x_{n,t+1}$ for each firm $n$. The ``lags” of a time series $x_t$ are considered to be all the past observations of $x_t$ , i.e $x_{t-1}$ , $x_{t-2}$, .... The semilogy of the mean squared error loss during training for every $10$ batches of $128$ samples is shown in Fig.\ref{Fig9-10} on the left.

The actual versus the predicted values from this model are
shown in Fig.\ref{Fig9-10} on the right. Here we randomly pick from the test set values of prices for some firm in the sample at some point in time. We plot these values with black and for each of them, we show with a blue color the predicted values from the model. The mean squared error loss is $0.06$.

We also report the results of the model when we use different
number of hidden layers and different number $m$ of lags of
$x_{n,t}$ to predict the next values of $x_{n,t+1}$ for firm $n$. These results are shown in Table \ref{Tab2}. From these results, we can see that the minimum value of the mean squared error is achieved when using $1$ hidden
layer, $x_t$ and $9$ lags of $x_t$ as input $(m = 10)$ to predict $x_{t+1}$.

\begin{table}
\centering
\vspace{2ex}
\begin{tabular}{|l| l |l| l| l|}\hline
m&$15$&$10$&$5$&$1$\\\hline 
$1$ Hidden Layer & $0.0623$ & \textbf{0.06}& $0.0601$& $0.0613$\\\hline
$2$ Hidden Layers & $0.0612$& $0.0718$ & $0.0630$& $0.0675$\\\hline
\end{tabular}
\caption{Mean Squared Error Comparison}
\label{Tab2}
\end{table}

\subsubsection{Non random split of train and test set}

\begin{figure}
\centering
\includegraphics[width=0.47\linewidth]{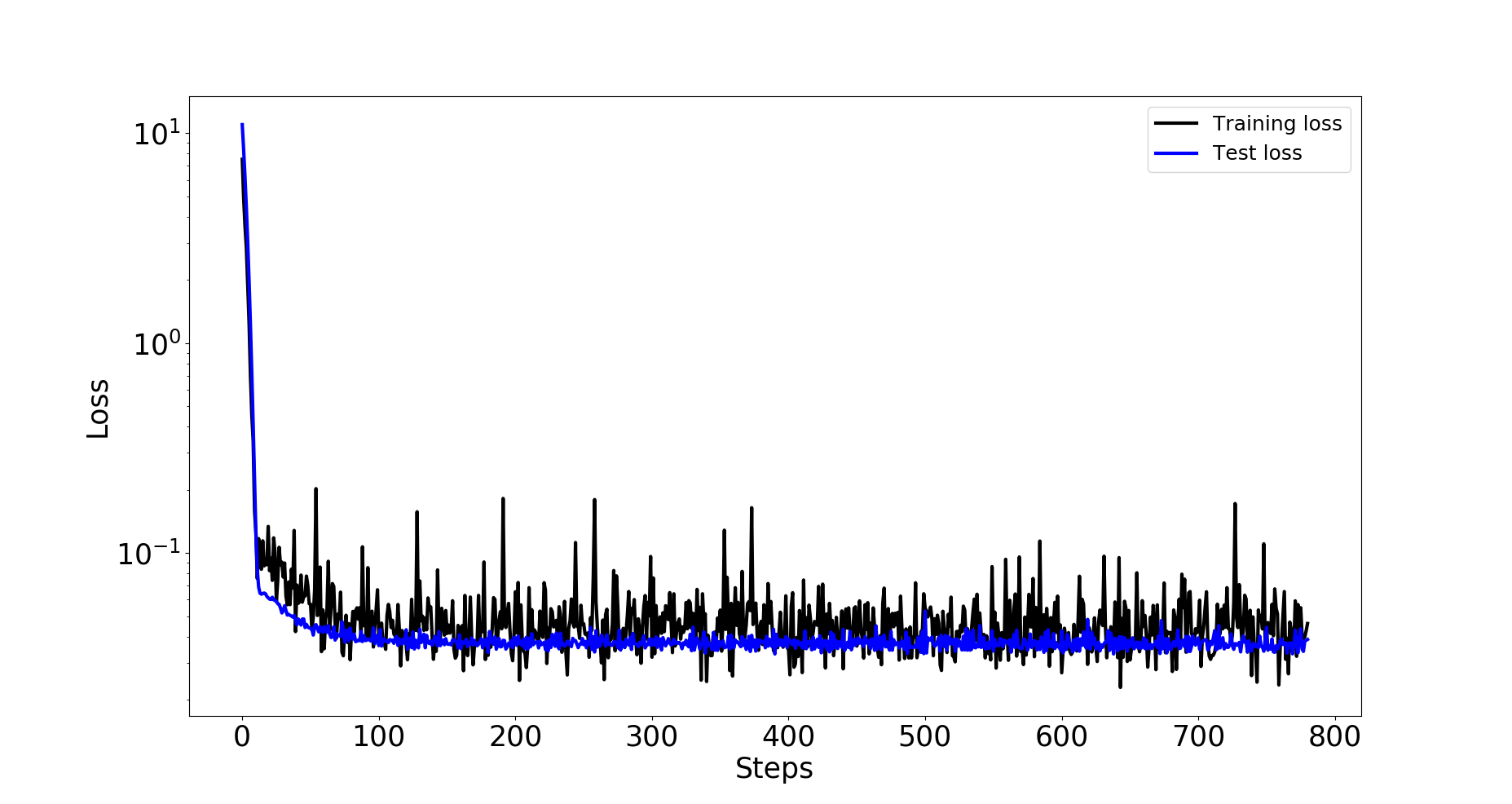}
\includegraphics[width=0.47\linewidth]{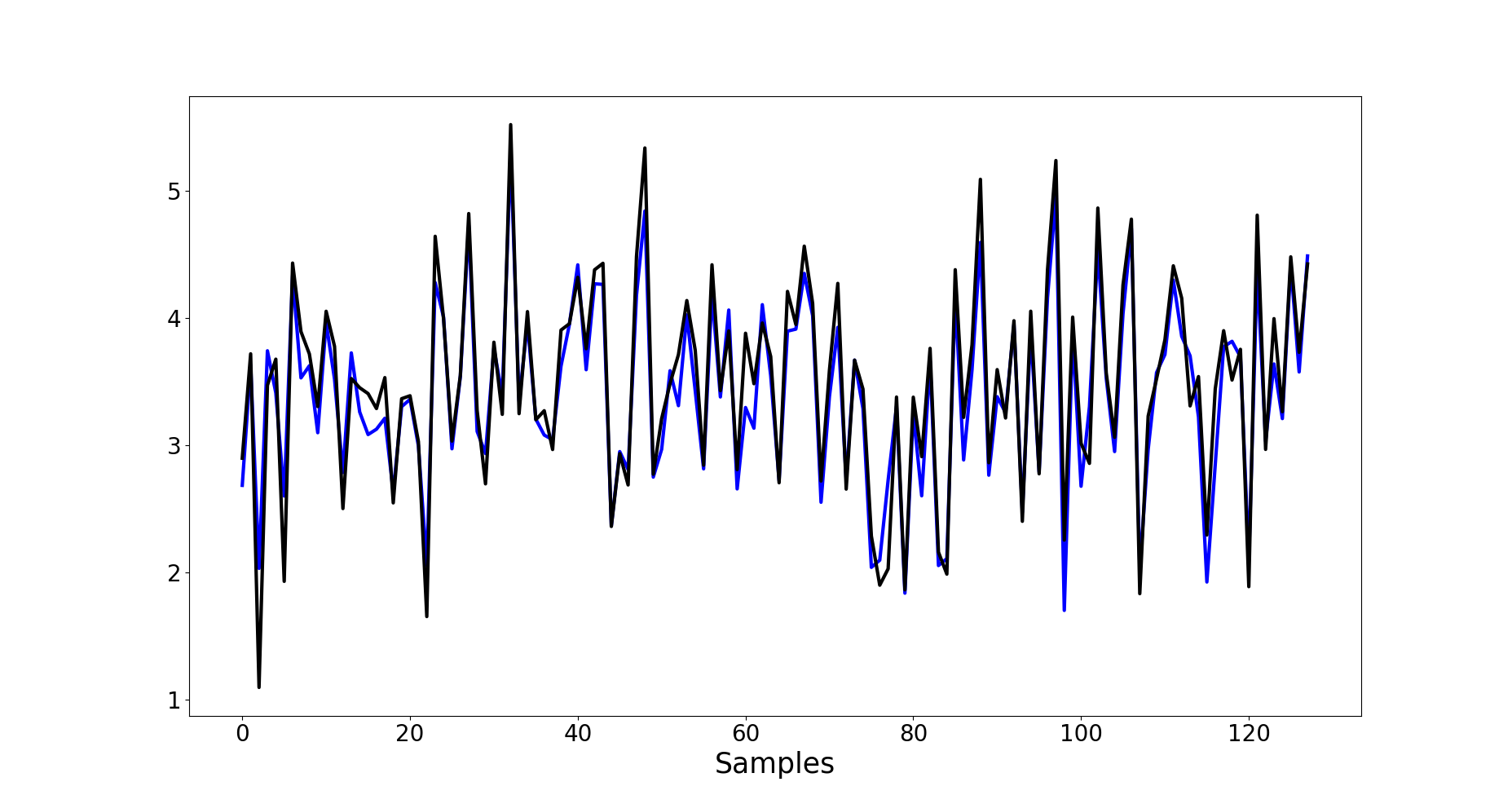}
\caption{Left: Semilogy of the mean squared error loss during training for MLP; Right: Predicted (blue) vs Actual (black) stock prices for random actual-predicted pairs in the test set for MLP with non random splitting}
\label{Fig11-12}
\end{figure}

For this section we use the variables again in their logarithmic form. Fig.\ref{Fig11-12} on the left shows the semilogy of the mean squared error loss during training for every $10$ batches of $128$ samples. Fig.\ref{Fig11-12} on the right shows the actual values (black line ) versus the predicted values (blue line) for some randomly picked observations in the sample.

Table \ref{Tab3} shows the comparison for different model specifications in terms of the mean squared error. The best specification seems to be the MLP with $m$ value of $10$ and $2$ hidden layers, where the error reached is $0.0337$. The error is lower compared to the model where the split is random because the test size here is smaller. In the random split, the test size consists of $33\%$ of the sample, while here it contains only observations in $2016$ and after. 

\begin{table}
\centering
\vspace{2ex}
\begin{tabular}{|l| l |l| l| l|}\hline
m&$15$&$10$&$5$&$1$\\\hline 
$1$ Hidden Layer & $0.0427$ & $0.041$& $0.0399$& $0.0437$\\\hline
$2$ Hidden Layers & $0.0401$& \textbf{0.033} & $0.0372$& $0.0348$\\\hline
\end{tabular}
\caption{Mean Squared Error Comparison for MLP with non random splitting}
\label{Tab3}
\end{table}

\section{A test for the relevance of network inputs}
\label{testsection}
In this section we formally test the null hypothesis that past prices do not affect present stock prices versus the alternative that they do. For this task, we use the results from \cite{white2001statistical}. The authors use a one hidden layer perceptron, with a continuous nonlinearity $\psi$ for the hidden layer and a linear activation function for the output layer. Formally, the output is given by:
\begin{equation}
f(x,w)=w_{00}+\sum_{j=1}^J w_{0j} \psi(\tilde{x}^Tw_{1j}),
\label{eq51}
\end{equation}
where $w \equiv (w_{00}, w_{01},...,w_{0h}, w_{11}^T,...,w_{1h}^T)^T$ are the weights, $x$ denotes the vector of inputs with $\tilde{x}=(1,x^T)^T$ and $h$ is the number of neurons in the hidden layer. We use 1 hidden layer with 16 neurons and a hyperbolic tangent function as $\psi()$. Learning is done to minimize the following:
\begin{equation}
\min_{w}N^{-1} \sum_{n=1}^N(Y_n-f(X_n,w))^2,
\end{equation}
where $N$ is the number of observations, $Y_n$ denotes the target and $f(\cdot, \cdot)$ is the function defined in Eq. \ref{eq51}. During training, we use a piecewise constant learning rate with boundaries at $600000$ and $1200000$ and values of the learning rate of $0.004$, $0.0005$ and $0.0001$ respectively for each interval. Observations before 2012 are used for the train set and observations after 2012 are used for the test set. 

With the appropriate restrictions discussed in \cite{white2001statistical}, it follows that a weight vector $\hat{w}_N$ exists and converges almost surely to $w^*$ which solves:
\begin{equation}
\min_{w} \lambda(w)=E([Y_n-f(X_n,w)]^2).
\end{equation}
The hypothesis we are interested to estimate is:
\begin{equation}
\partial f(x,w^*)/ \partial x_i =0 , \quad i \in I_0,
\end{equation}
where $I_0$ is the set of indexes specifying the inputs whose relevance we are interested in. The authors consider the following statistics:
\begin{equation}
 \hat{m} = N^{-1}\sum_{n=1}^N \sum_{i \in I_0} f_i(X_n,\hat{w}_N)^2,
\end{equation}
where $f_i(\cdot, w)$ denotes $\partial f(\cdot, w)/ \partial x_i$. This statistics is $0$ if and only if $f_i(X_n,\hat{w}_N)$ is $0$. They show that the statistics has an asymptotic $\chi2$ mixture distribution. Since the computation of a critical value is complicated, the authors use the bootstrap method with a bootstrap statistics given by:
\begin{equation}
\bar{\mathcal{B}}^*_N=\sum_{n=1}^Nm(X_n,\hat{w}^*_N)-\sum_{n=1}^N m(X_n,\hat{w}_N) - \sum_{n=1}^N \bigtriangledown ^T m(X_n,\hat{w}_N)(\hat{w}^*_N - \hat{w}_N).
\end{equation}
Here $\hat{w}^*_N$ are the weights obtained from the bootstrap sample and $m(X_n,\hat{w}_N)=\sum_{i \in I_0} f_i(X_n,\hat{w}_N)^2$. The bootstrap method consists of the following steps:
\begin{enumerate}
\item Use the original sample solve the minimization problem to get $\hat{w}_N$.
\item Draw a sample $\{Z_1^*,...,Z_N^*    \}$ with replacement from the original sample and compute resampled weights $\hat{w}_N^*$ by solving the minimization problem with the new sample. 
\item Compute the bootstrap statistics, $\bar{\mathcal{B}}^*_N/N$.
\item Repeat steps 2 and 3 $v$ times (e.g $v=100$ or $v=1000$). 
\item Compute the $p$-value. 
\end{enumerate}
We use $1000$ bootstrap samples for each model. The weights $\hat{w}^*_N$ for each of the bootstrap sample are learned by setting the initial values of the weights at $\hat{w}_N$.

Following the methods described in \cite{white2001statistical}, one should train separately for each firm, for each frequency (daily, weekly, monthly),  and for different time periods. Due to the computational intensity of experimenting with $439$ firms, $3$ different frequencies and different time periods, we only perform the test on the top $5$ firms with monthly frequency and a time period from 1980 to 2017. Thus, our hypothesis testing is now restricted to "are past prices of those 5 firms relevant in predicting the current prices?". The firms used are the top 5 companies in terms of Market Capitalization from NASDAQ (as of 2017): Apple, Inc. (AAPL), Comcast Corporation (CMCSA), Intel Corporation (INTC), Microsoft Corporation (MSFT), PepsiCo, Inc. (PEP), monthly frequency and the 3 models as in Eq. \ref{Eq18}, \ref{Eq19} and \ref{Eq20}.

The $p$-value is equal to the proportion of bootstrap samples that give a higher statistics than the original one. For example, if the number of samples exceeding the original statistics is $k$ and the total number of bootstrap statistics is b, the $p$-value is calculated as $\frac{k}{b}$. In this work we use a ratio of $\frac{k+1}{b+1}$ to provide more robust estimates. Also, we use percentage price changes, $\bigtriangleup \log P_t$ and for a given firm we estimate the three of the following models:
\begin{equation}
\bigtriangleup \log P_t=f(\bigtriangleup \log P_{t-1},\bigtriangleup \log P_{t-2},\bigtriangleup \log P_{t-3},\bigtriangleup \log P_{t-4},\bigtriangleup \log P_{t-5},\omega)+\epsilon_t
\label{Eq20}
\end{equation}
\begin{equation}
\bigtriangleup \log P_t=f(\bigtriangleup \log P_{t-1},\bigtriangleup \log P_{t-2},\bigtriangleup \log P_{t-3},\omega)+\epsilon_t
\label{Eq19}
\end{equation}
\begin{equation}
\bigtriangleup \log P_t=f(\bigtriangleup \log P_{t-1},\omega)+\epsilon_t.
\label{Eq18}
\end{equation}

\begin{figure}[!htb]
\centering
\includegraphics[width=0.3\linewidth]{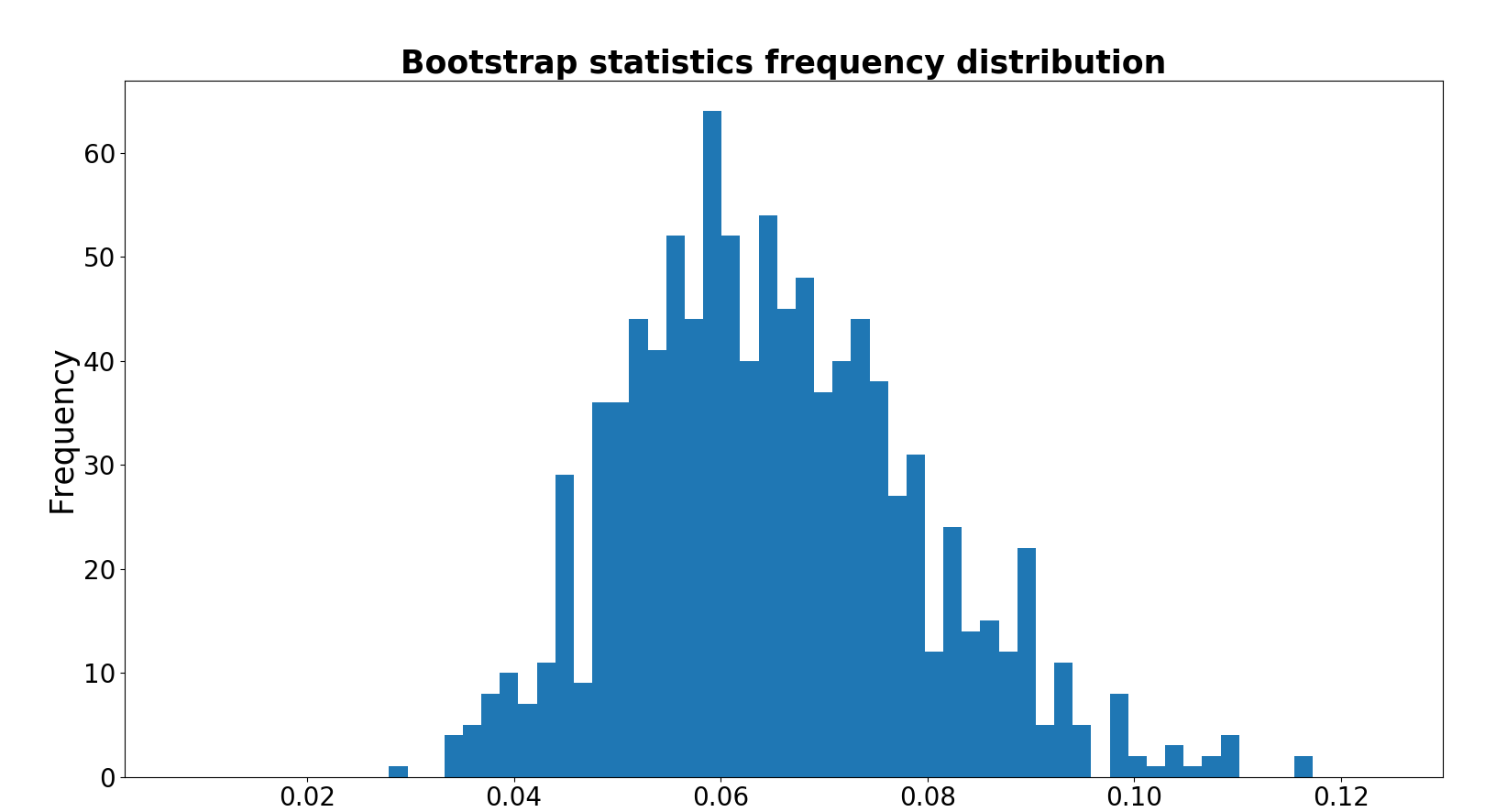}
\includegraphics[width=0.3\linewidth]{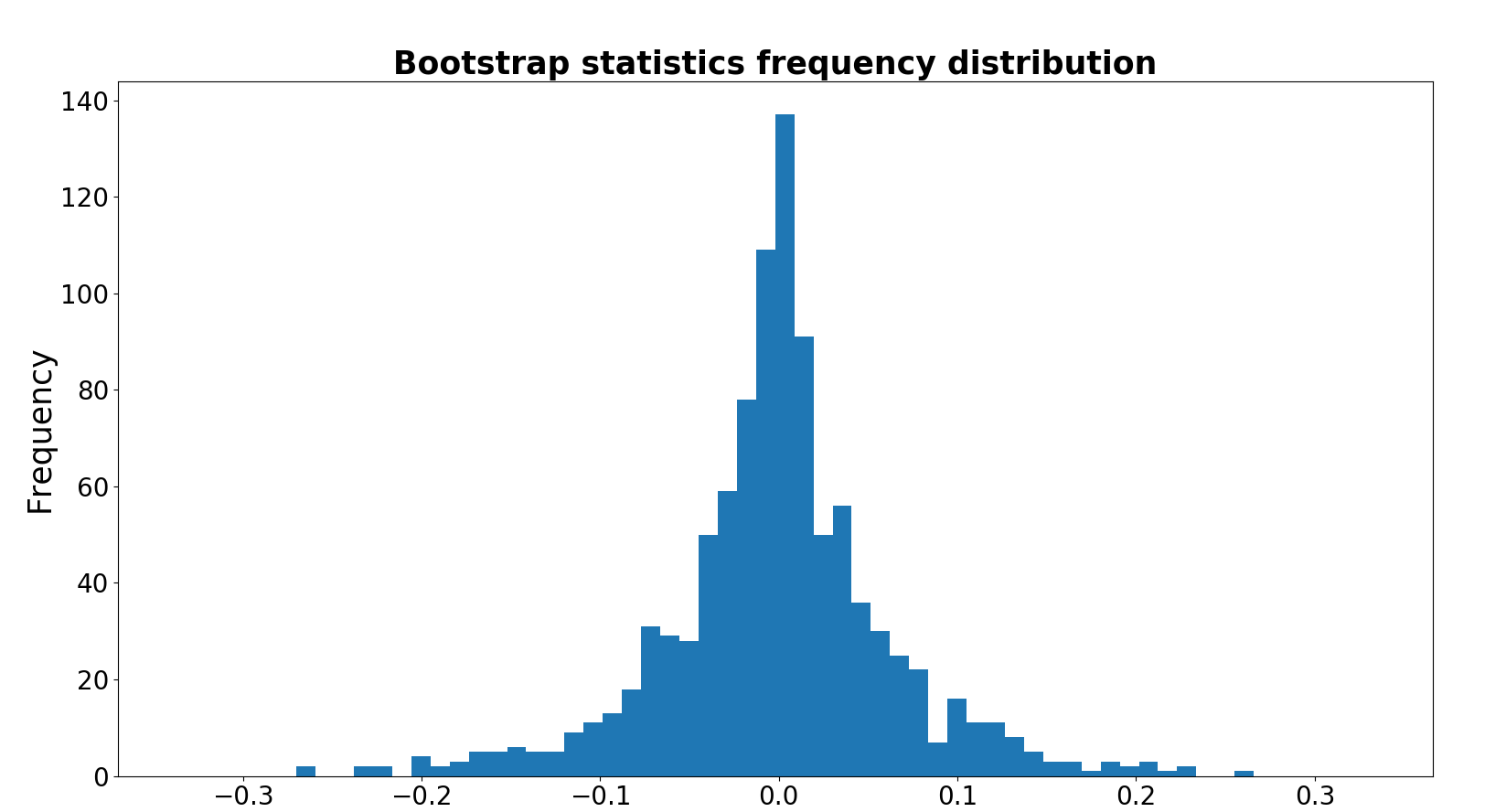}
\includegraphics[width=0.3\linewidth]{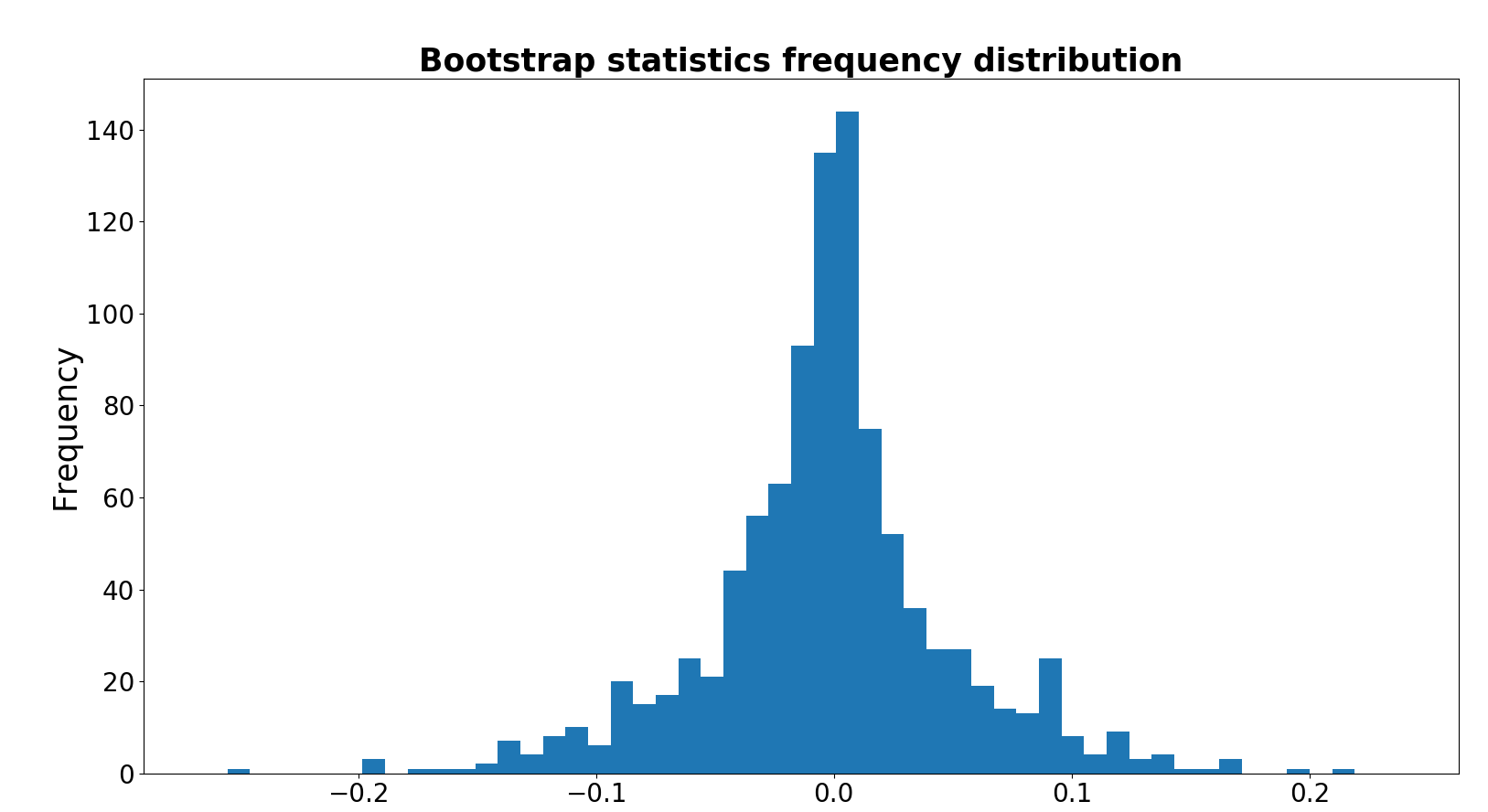}\\
\includegraphics[width=0.3\linewidth]{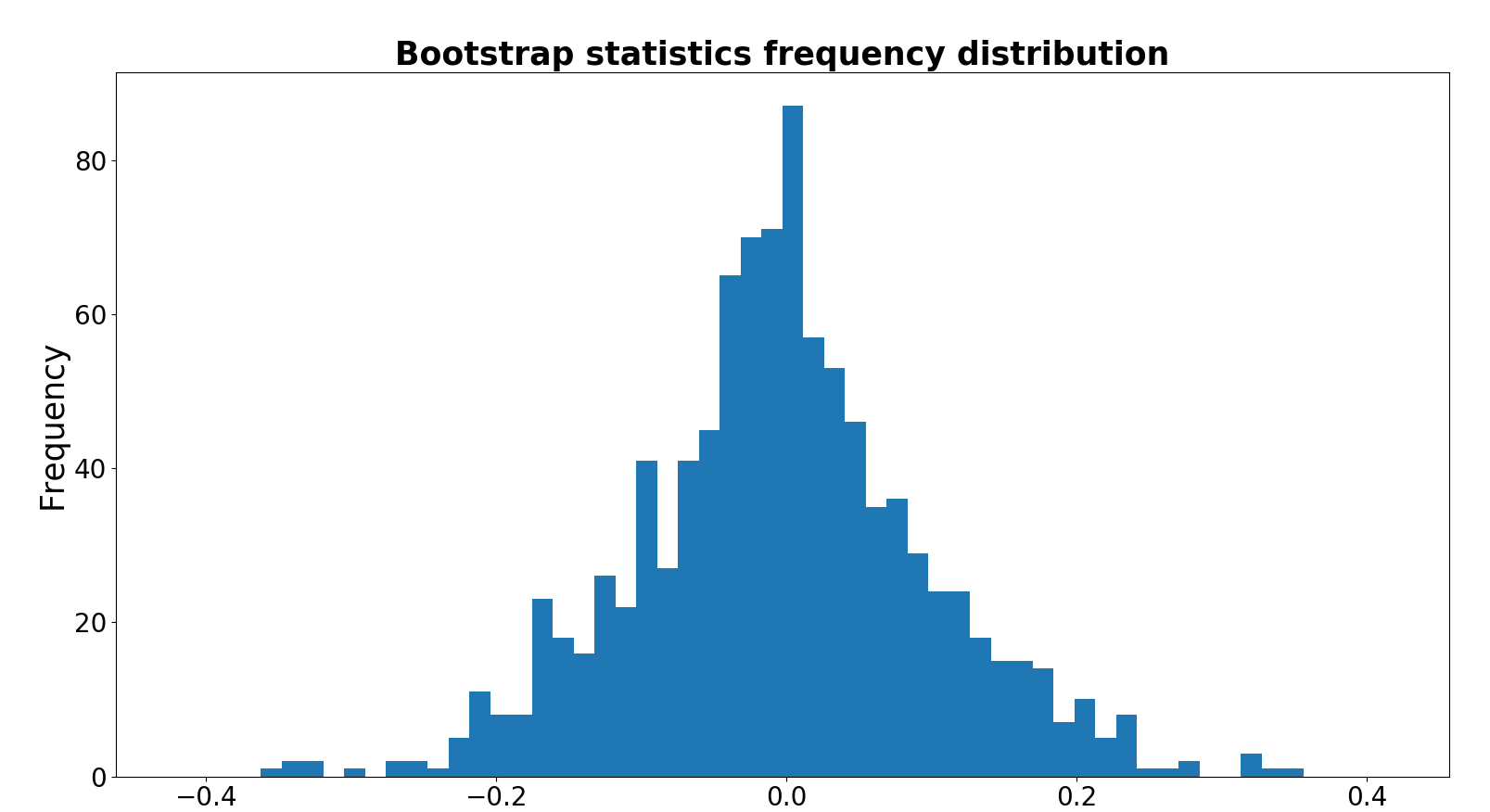}
\includegraphics[width=0.3\linewidth]{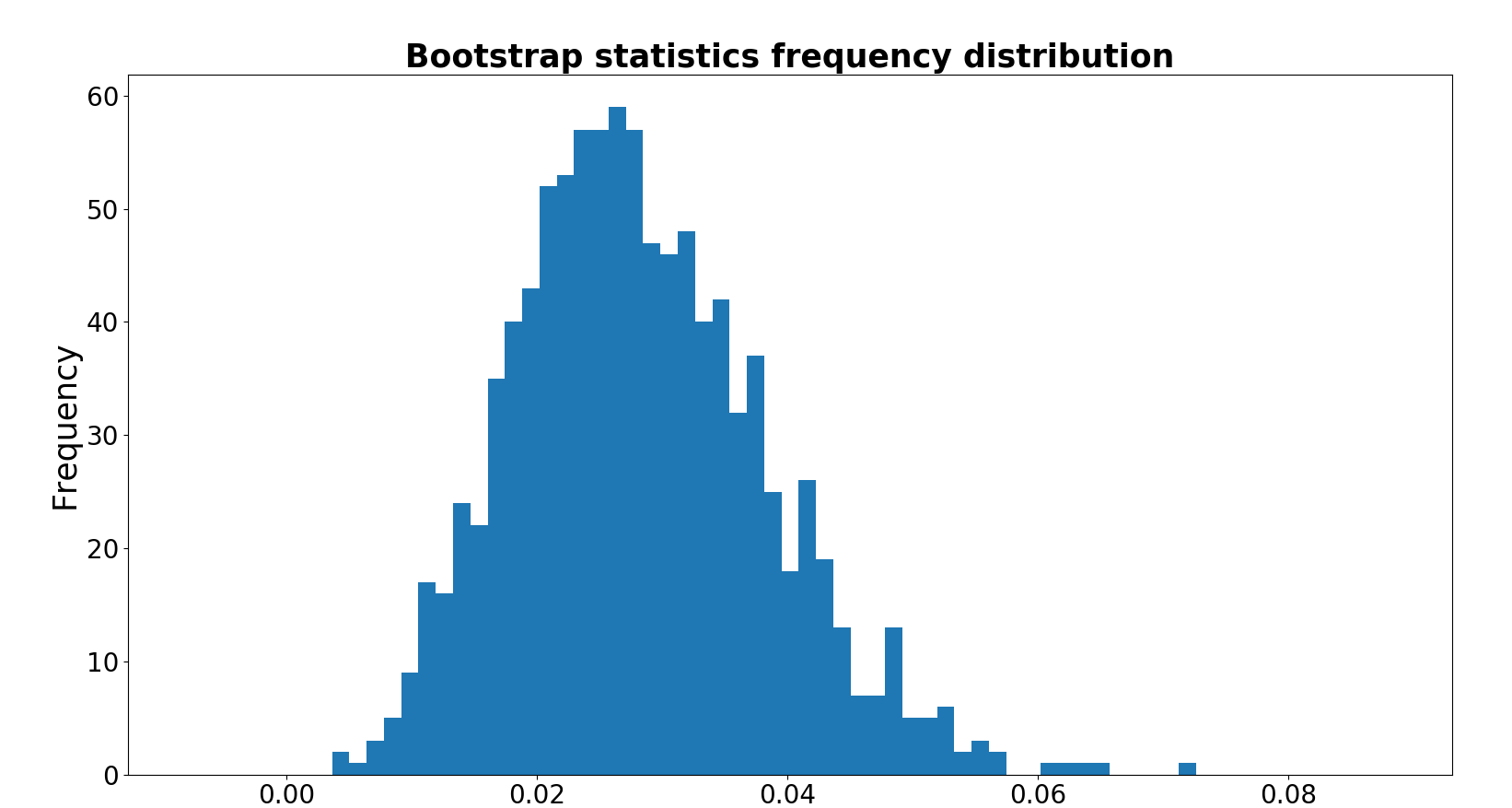}
\caption{Bootstrap Statistics Distribution for AAPL, CMCSA, MSFT, INTC and PEP}
\label{Fig13}
\end{figure}

\begin{figure}[!htb]
\centering
\includegraphics[width=0.5\linewidth]{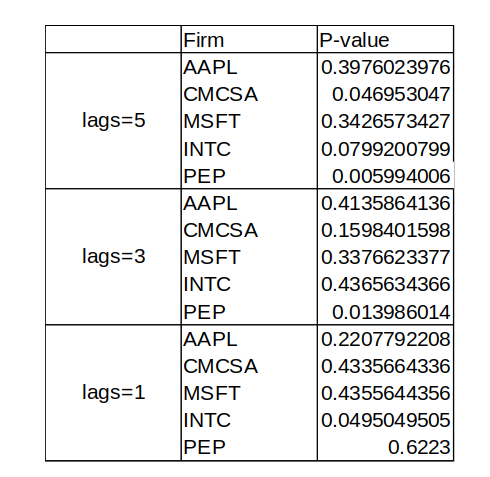}
\caption{p-values for each individual model}
\label{Fig14}
\end{figure}

Then we have to test the null hypothesis with multiple models, for different lags used and different firms. We need an upper bound on the $p$-value for the joint null hypothesis that in none of a particular group of models do the lagged changes matter. For this, we use the Bonferroni inequality for multiple hypothesis. With the Bonferroni method, we reject the null at the $\alpha$ level if $P_{(1)} \leq \alpha/m$, where $m$ is the number of models and  $P_{(1)} $ is the smallest $p$-value from all the $m$ models. So the Bonferroni $p$-value bound is $\alpha=mP_{(1)}$. One could use tighter bounds by a modification of the Bonferroni inequality as in \cite{hochberg1988sharper}.

Fig. \ref{Fig13} shows the bootstrap statistics for all the firms, trained with the model with 5 lags. The statistics does not seem to be normally distributed nevertheless the distribution seems approximately symmetric, with the exception of AAPL and PEP. The p-values for all $15$ models we estimate are given in Fig. \ref{Fig14}. With a value of $P_{(1)}$ equal to $0.0059$, $\alpha=10\%$ and $m=15$, we can reject the null hypothesis that in none of the models do lagged changes matter. 

\section{Conclusions}
This work questions the efficient-market hypothesis using DNNs. For this purpose, we use RNN and MLP to forecast next quarter’s stock movements using historical stock prices. We train the MLP with a random and a non random train and test split. We also conduct a formal statistical test for the null hypothesis that past prices do not affect current prices versus the alternative that they do. The results reject the null hypothesis that in none of the models that we estimate, do the lagged prices matter. These results contradict the efficient-market hypothesis, in line with the work of \cite{basu1977investment}, \cite{rosenberg1985persuasive} and others. This encourages the use of better models, different specifications and different processing of the data to predict more accurately the future stock movements. 

\newpage
\bibliography{ref}
\bibliographystyle{apalike}
\clearpage
\end{document}